\documentclass[11pt,tightenlines]{revtex4}
\usepackage{amsfonts}
\usepackage{amsmath}
\usepackage{txfonts}
\usepackage{graphicx}
\usepackage{amssymb}
\usepackage{color}
\usepackage{hyperref}

\usepackage{sidecap}

\begin{document}
\title{Persistent memories in transient networks}
\author{Andrey Babichev$^{1,2}$ and Yuri Dabaghian$^{1,2*}$}
\affiliation{$^1$Jan and Dan Duncan Neurological Research Institute, Baylor College of Medicine,
Houston, TX 77030, \\
$^2$Department of Computational and Applied Mathematics, 
Rice University, Houston, TX 77005\\
$^{*}$e-mail: dabaghia@bcm.edu}
\vspace{17 mm}
\date{\today}
\vspace{17 mm}
\begin{abstract}
Spatial awareness in mammals is based on an internalized representation of the environment, encoded by large 
networks of spiking neurons. While such representations can last for a long time, the underlying neuronal network 
is transient: neuronal cells die every day, synaptic connections appear and disappear, the networks constantly 
change their architecture due to various forms of synaptic and structural plasticity. How can a network with a 
dynamic architecture encode a stable map of space? We address this question using a physiological model of a 
``flickering'' neuronal network and demonstrate that it can maintain a robust topological representation of space.
\end{abstract}
\maketitle

\newpage

\section{Introduction}
\label{section:intro}
It is believed that mammalian ability to navigate, to escape from predators, to find its nest, to plan hunting strategies 
and so forth, is based on an internalized neuronal representation of space---a cognitive map of the environment 
\cite{OKeefe,Tolman}. Neurophysiologically, this map is encoded via timed sequences of quick electrical discharges 
of the neurons---neuronal spikes---in various parts of the brain. A particularly important role in producing this map 
is played by the hippocampus---one of the oldest parts of the mammalian brain in evolutionary terms. 
The spike times of the hippocampal principal neurons---so called place cells---are defined by the animal's spatial location: in 
rodents, each place cell fires when the animal visits a specific location---this cell's place field \cite{Tolman} (Fig.~\ref{PFs}A). 
Why a given place cell fires only when an animal is ``here'' rather than ``there'' is still a mystery, and how the ensemble 
of place cells forms a hippocampal map of the environment is equally mysterious. In particular, it remains largely 
unknown how the spike trains produced by the place cells are processed downstream from the hippocampus. 

Experimental evidence points out that groups of coactive place cells form functionally interconnected ``assemblies'' 
\cite{Harris, Buzsaki} that together drive their respective ``read-classifier'' or ``readout'' neurons in the downstream 
networks (Fig.~\ref{PFs}B). Since coactivity of the place cells marks the overlap of their respective place fields (Fig.~\ref{PFs}C), 
the activity of a readout neuron actualizes a connectivity relationship between the regions encoded by the individual 
place cells. This suggests that the hippocampus provides a qualitative representation of space, based on connectivity, 
adjacency and containment relationships, i.e., that the hippocampal cognitive map is topological in nature---a hypothesis 
that receives an increasing amount of experimental support\cite{Gothard,Leutgeb,Alvernhe,Poucet,eLife}.

In \cite{PLoS,Arai} we proposed a theoretical model that showed that hippocampal cell assembly network can indeed 
capture the spatial connectivity of the environment in a biologically plausible time, given that the place cells operate 
within biological parameters of firing rate and place field size. However, the approach of \cite{PLoS,Arai} was based 
on analyses of the ever-growing pool of spike trains, i.e., it ignored that, in physiological networks, the connectivity 
information may not only accumulate, but also decay. In particular, the physiological cell assemblies may not only form, 
but also disband as a result of deterioration of synaptic connections \cite{Wang}, then reappear as a result of subsequent 
learning, then disband again and so forth. Electrophysiological studies suggest that the lifetime of the cell assemblies ranges 
between tens of milliseconds to minutes or longer \cite{Harris,Buzsaki,Bi,Magee}, whereas spatial memories in rats can last 
for days and months \cite{Meck,Clayton,Brown}. How can the large-scale spatial representation of the environment be stable 
if the neuronal stratum that computes this representation changes on a much faster timescale? Below we discuss a model of 
a transient hippocampal network and use methods of Algebraic Topology to demonstrate that the topological characteristics of 
the large-scale spatial representation of the environment encoded by this network can remain stable. 

\section{Model}
\label{model}

\begin{figure} 
\includegraphics[scale=0.84]{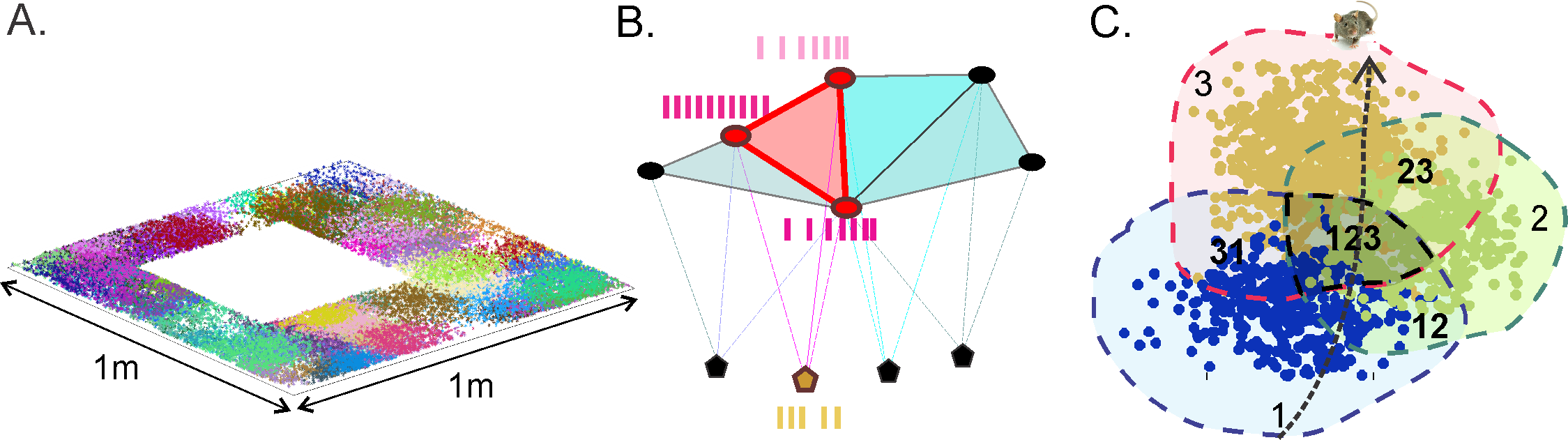}
\caption{\label{PFs} \textbf{Place cells and cell assemblies}. (\textbf{A}). Simulated place fields in a small planar environment 
containing a square hole: dots of different color mark the locations where the corresponding place cells produced spikes. The 
dots of a particular color form spatial clusters---the place fields. (\textbf{B}). Schematic representation of four overlapping 
third-order cell assemblies by four third order simplexes (triangles). The vertexes (shown by small circles) represent place cells 
synaptically connected to their respective readout neurons (pentagons). A cell assembly $\sigma$ ``ignites'' (red triangle) when 
all of its place cells become simultaneously active, thus eliciting a response from the readout neuron $n_\sigma$ (active cells have 
colored centers). (\textbf{C}). A coactivity of the place cells $c_1$, $c_2$ and $c_3$ occurs when the rat crosses through the 
overlap of the corresponding place fields---the domain 123.} 
\end{figure}

The way place fields cover an environment $\mathcal{E}$ (Fig.~\ref{PFs}A) calls to mind a basic theorem of algebraic 
topology due to P. Alexandrov \cite{Alexandroff} and E. \v{C}ech \cite{Cech}: it is possible to reconstruct the topology of 
a space $X$ from the pattern of overlaps between the regions that cover it. The model proposed in \cite{Curto,PLoS,Arai} 
is based on the observation that this theorem can not only be applied to the place fields themselves, but also implemented 
via spiking signals of cells that represent these place fields. In this approach, groups of coactive place cells, $c_1$, $c_2$, 
..., $c_n$, are viewed as abstract simplexes, $\sigma = [c_1, c_2, ...,c_n ]$, which together form a simplicial ``cell assembly 
complex,'' $\mathcal{T}_{CA}$ (for definitions and details see  \cite{Hatcher, Aleksandrov} and Methods in \cite{PLoS}). 

This construction provides a connection between the local (cellular) and the global (system-level) scales: the individual cell 
assemblies, just like simplexes, provide local information about the space, but together, as a neuronal ensemble, they represent 
space as whole---as the simplicial complex. Thus, $\mathcal{T}_{CA}$ provides a schematic representation of the cell assembly 
network and of its rewiring dynamics: formation of new place cell assemblies and disbanding of some old ones are represented, 
respectively, by the appearance and disappearance of their counterpart (maximal) simplexes in $\mathcal{T}_{CA}$. 
%%%%%%%%%%%%%%%%%%%%%%%%%%%%%%%%%%
%%%%%%%%%%%%%%%%%%%%%%%%%%%%%%%%%%

\begin{SCfigure}
\centering
\includegraphics[scale=0.75]{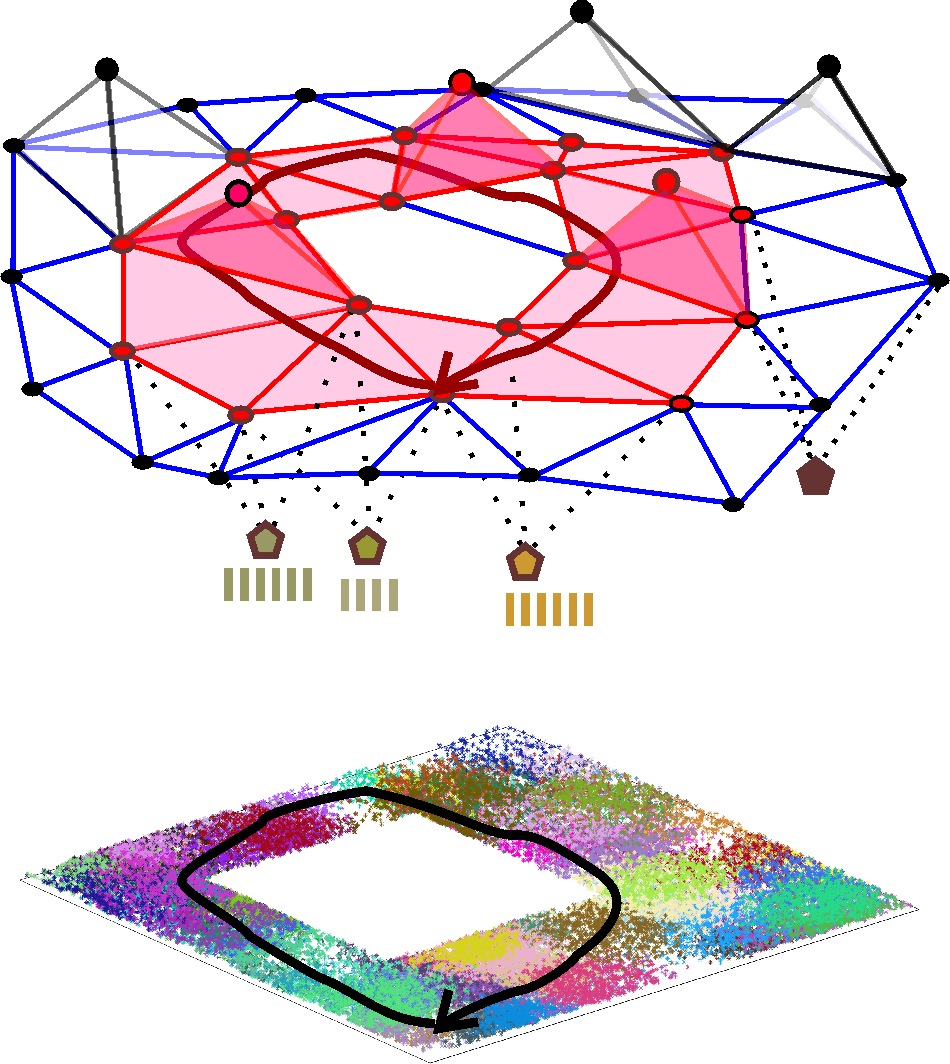}
\caption{\label{Schem} \textbf{Scehamtic representation of place cell coactivity in the cell assembly complex $\mathcal{T}_{CA}$}. 
The topological structure of $\mathcal{T}_{CA}$, induced from the place field layout in $\mathcal{E}$, provides a topological 
representation of $\mathcal{E}$. For example, the hole in the middle of $\mathcal{T}_{CA}$, which produces non-contractible 
simplicial paths (topological loops) corresponds to the central hole in the environment $\mathcal{E}$, which produces non-contractible 
navigation paths $\gamma$ (e.g., a sample path is shown at the bottom). As the animal travels along $\gamma$, the hippocampal place 
cell assemblies ignite in a certain order, which corresponds to a chain of maximal simplexes in $\mathcal{T}_{CA}$---a simplicial 
path $\Gamma$, shown by red triangles and tetrahedrons.} 
%\end{figure} 
\end{SCfigure}

On the other hand, the cell assembly complex $\mathcal{T}_{CA}$ provides semantics for describing the global spatial memory 
map in topological terms  \cite{Schemas}. For example, a sequence of cell assemblies ignited along a path $\gamma$ navigated 
by the rat corresponds to a chain of simplexes $\Gamma \in \mathcal{T}_{CA}$---a ``simplicial path'' (Fig.~\ref{Schem}). The 
structure of $\mathcal{T}_{CA}$ allows to establish certain qualitative properties of the simplicial paths and to relate them to the 
properties of the physical paths in $\mathcal{E}$. For example,  a simplicial path that closes onto itself in $\mathcal{T}_{CA}$ may 
represent a closed physical path, a pair of topologically equivalent simplicial paths $\Gamma_1 \sim \Gamma_2$ may represent 
physical paths $\gamma_1$ and $\gamma_2$ that can be deformed into one another, a hole in $\mathcal{T}_{CA}$ may represent 
a physical obstacle and so forth. However, establishing these correspondences requires learning: as the animal 
begins to explore the environment, only a few place cells would have time to (co)activate, and only a few cell assemblies would 
have time to form; as a result, a newly developing complex $\mathcal{T}_{CA}$ consists of only a few maximal simplexes and 
contains many holes, some of which correspond to physical obstacles or to regions that have not yet been visited by the animal, 
and others are ``spurious'', i.e., reflecting transient information about place cell coactivity. As shown in \cite{PLoS,Arai}, the 
spurious holes tend to disappear as more spatial information accumulates.

In \cite{Babichev} we suggested two methods for constructing the cell assembly network and hence producing the simplicial cell 
assembly complex $\mathcal{T}_{CA}$ that represents this network, by selecting the most frequent combinations of spiking 
place cells. The key observation of  \cite{Babichev} was that in bounded environments, the coactive cell combinations eventually 
become repetitive, and it is therefore possible to identify the cell assemblies from the most frequent coactivity patterns. However, 
the frequency of a given cell assembly's activation, $f_{\sigma}$, was evaluated from the total number of its appearances over 
the entire navigation period, and then the selected cell assemblies were presumed to have existed for as long as the navigation 
continued. In other words, the cell assemblies were ``perennial'' by construction. In the following, we model a transient hippocampal 
network by limiting the time during which a cell assembly $\sigma$ can form to a smaller time interval---``a memory window'' 
$W^{(\sigma)}$. Physiologically, $W^{(\sigma)}$ defines the period during which readout neuron $n_{\sigma}$ can identify the 
combination $\sigma$ of coactive place cells, connect to it synaptically, retain these connections and respond to subsequent 
ignitions of $\sigma$.

To simplify the approach, we consider the case in which the entire ensemble of readout neurons is characterized by a single 
parameter $W^{(\sigma)} = W$, and proceed as follows. First, we identify the cell assemblies that emerge within the first 
$W$-period after the onset of the navigation, $W_1$, and then repeat the algorithm for the subsequent windows $W_2$, 
$W_3$,..., obtained by shifting the $W_1$ over small time steps. As a result, a cell assembly $\sigma$ that was identified in 
the window $W_{t_1}$, based on the local place cell coactivity rate $f_{\sigma}(W_{t_1})$, may disappear at a certain step 
$W_{t_2}$, then reappear in a later window, $W_{t_3}$, disappear again and so forth. The ensemble of appearing and 
disappearing cell assemblies can then be schematically represented by a ``flickering'' simplicial complex, $\mathcal{F}_{W}$, 
with appearing and disappearing maximal simplexes. Our task is to study the net topological properties of $\mathcal{F}_{W}$, 
e.g., whether the lifetimes of its topological loops can be longer than the lifetimes of its typical maximal simplex.

%%%%%%%%%%%%%%%%%%%%%%%%%%%%%%%%%%%
%%%%%%%%%%%%%%%%%%%%%%%%%%%%%%%%%%%

\begin{figure} 
\includegraphics[scale=0.87]{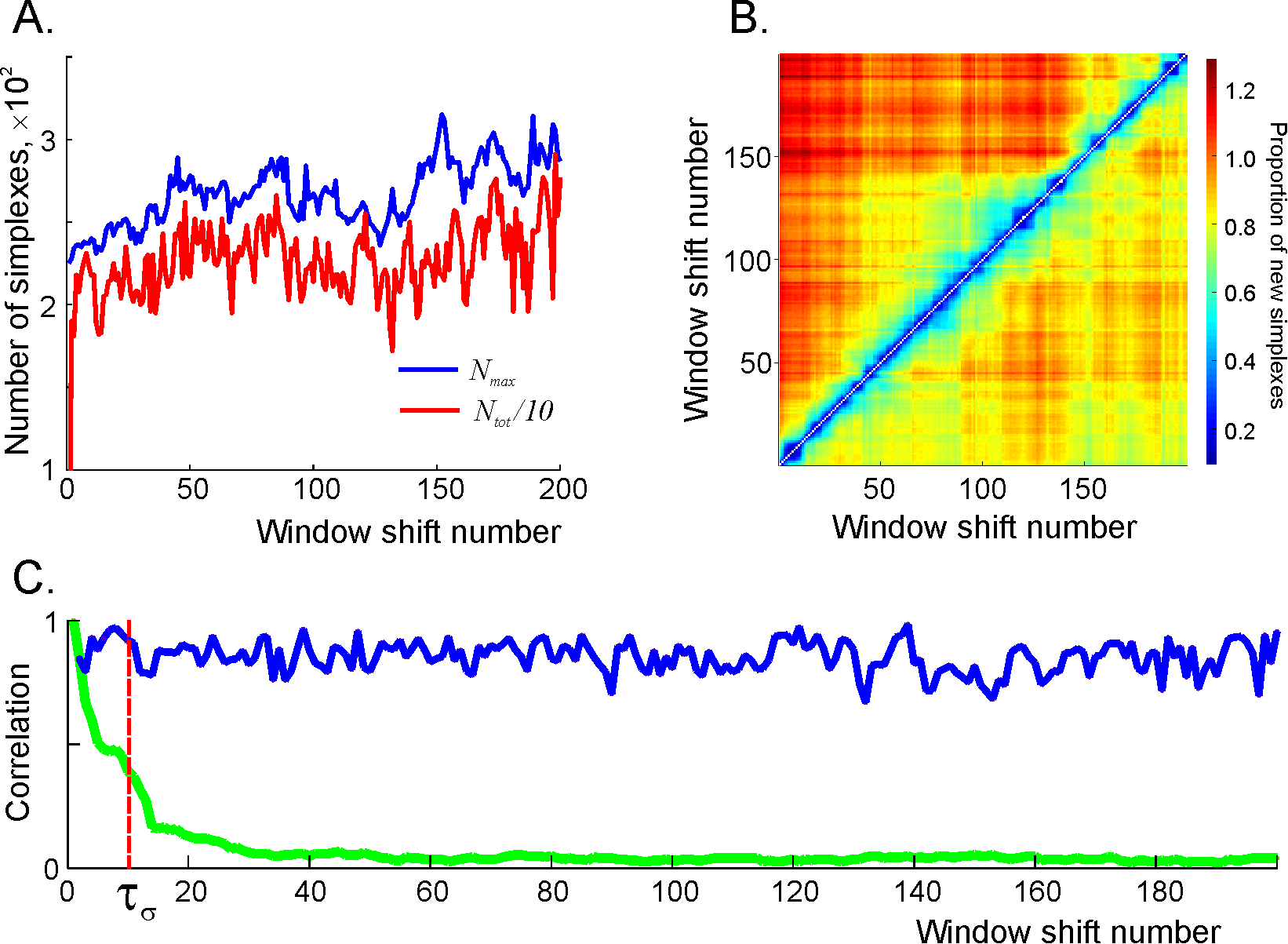}
\caption{\label{Flick} \textbf{Time evolution of the cell assembly complex}. (\textbf{A}) The number of maximal simplexes 
(blue trace) and the total number of all simplexes, scaled down by a factor of 10 (red) in the flickering simplicial complex 
$\mathcal{F}_{W}$. The memory window spans over $W = 5$ minutes, shifting over $\Delta t = 2.5$ secs at a time. Shown 
are first 200 shifts. (\textbf{B}) The matrix of similarity coefficients, $r_{ij}$, between the pairs of flickering cell assembly 
complexes $\mathcal{F}_{W}(t_i)$ and $\mathcal{F}_{W}(t_j)$ defined as the proportion of the maximal simplexes contained 
in $\mathcal{F}_{W}(t_i)$ but not in $\mathcal{F}_{W}(t_j)$. For close moments $t_i$ and $t_j$, the differences are small, 
but as time separation grows, the differences becomes larger. (\textbf{C}) At every moment of time $t_i$, the blue line shows 
the fraction of the maximal simplexes that were also present at the previous moment, $t_{i-1}$. The green line shows the 
fraction of the original maximal simplexes (from $\mathcal{F}_{W}(t_1))$ remaining in $\mathcal{F}_{W}(t_i)$. 
The population of simplexes changes entirely (by about $0.95\%$) in about 60 steps. The dashed line marks the mean simplex 
decay rate, $\tau_{\sigma} = 10$ window shifts (about 25 seconds).} 
\end{figure} 
%%%%%%%%%%%%%%%%%%%%%%%%%%%%%%%%%%%

\section{Results}
\label{section:results}

We studied the dynamics of the flickering complex $\mathcal{F}_{W}$ and the topological information encoded in it, as a 
function of the discrete time $t_n = n\Delta t$, $\Delta t = 2.5$ secs, for the memory window width $W = 5$ minutes. The 
results produced by a typical neuronal ensemble containing $N_c$ = 300 simulated place cells are shown on Fig.~\ref{Flick}. 
First, we observe that the fluctuations of the number of the maximal simplexes in $\mathcal{F}_{W}$, $N_{\sigma}$, remain 
confined within about $20\%$ from the mean, $220 \leq N_{\sigma} \leq 300$, so that the simulated hippocampal network 
contains about 260 fluctuating cell assemblies on average (Fig.~\ref{Flick}A). However, while the size of the flickering complex 
remains bounded, the pool of the maximal simplexes changes significantly: as temporal separation $\Delta_{ij} = | t_i - t_j |$ 
between the memory windows increases, the corresponding complexes $\mathcal{F}_{W}(t_i)$ and $\mathcal{F}_{W}(t_j)$ 
become more and more dissimilar (Fig.~\ref{Flick}B). After about 50 time steps ($\sim$ 120 seconds), the set of simplexes in 
$\mathcal{F}_{W}$ is essentially renewed, which implies that the cell assembly network, as described by the model, is completely 
rewired (Fig.~\ref{Flick}C). 

A typical maximal simplex lasts on average about 10 discrete time steps ($\tau_{\sigma}\approx 25$ seconds), which is close 
to the range of values established experimentally \cite{Buzsaki}. Such rapid rate of the simplex renewal in the flickering cell 
assembly complex allows us to address our main question: what is the dynamics of the topological characteristics of 
$\mathcal{F}_{W}$ and how do they correspond to the topology of the environment?

The topology of the environment can be described in terms of the Betti numbers---roughly speaking, a Betti number $b_n$ 
defines the number of $n$-dimensional topological loops in $\mathcal{F}_{W}$ (i.e., closed surfaces counted up to topological 
equivalence). In the case of the environment shown on Fig.~\ref{PFs}A, the Betti numbers are as follows: $b_0 = 1$ (i.e., 
the environment is connected), $b_1 = 1$ (i.e., there is one $1D$ topological loop that encircles the hole in the middle), while 
$b_{n > 1} = 0$ (no topological loops in higher dimensions). Using the methods of Persistent Homology 
\cite{Ghrist,Zomorodian,Edelsbrunner}, we evaluated these numbers for the flickering cell assembly complex for the sequence 
of windows, and found that $\mathcal{F}_{W}$ does, in fact, reliably capture the topological properties of the environment 
over long periods, which significantly exceed both the simplexes' lifetimes, $\tau_{\sigma}$, and the width of the memory 
window, $W$. 

\begin{figure} 
\includegraphics[scale=0.84]{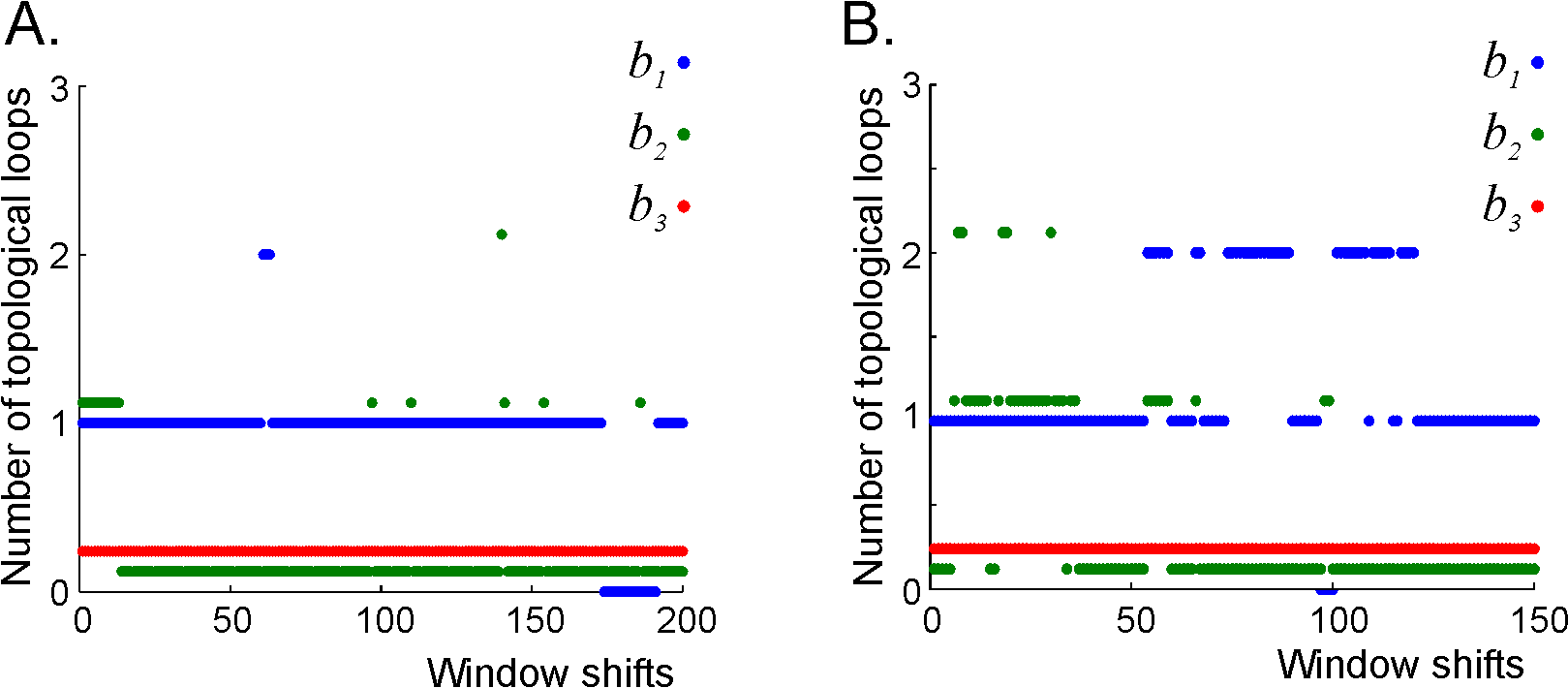}
\caption{\label{Betti} \textbf{ Stability of the topological information.} (\textbf{A}). The low-dimensional Betti numbers, $b_1$, 
$b_2$, $b_3$ as a function of the discrete time, computed using 5 minute wide memory window $W$. The 0th Betti number 
($b_0 = 1$ at all times) is not shown. (\textbf{B}). As the memory window decreases to $W = 3$ minutes, the frequency and 
the number of the topological fluctuations increases: the flickering complex repeatedly produces an extra (spurious) $1D$ 
topological loop and up to two spurious bubbles.} 
\end{figure}

As illustrated on Fig.~\ref{Betti}A, the first and the second Betti numbers almost never deviate from their respective physical 
values ($b_1 = 1$ and $b_2 = 0$). The occasional changes of $b_1$ and $b_2$ can be viewed as short-time ``topological 
fluctuations'' in the hippocampal map. For example, an occurrence of $b_1 = 2$ value indicates the appearance of an extra 
(non-physical) $1D$ loop, and at the moments when $b_1 = 0$, all $1D$ loops in $\mathcal{F}_{W}$ are contractible, i.e., 
the central hole (Fig.~\ref{PFs}A) is not represented in the hippocampal map. 
Similarly, the moments when $b_2 \neq 0$ indicate times when the flickering complex $\mathcal{F}_{W}$ contains non-physical, 
non-contractible $2D$ bubbles (one can speculate about the cognitive effects that these fluctuations may produce). The 0-th 
and the third Betti number always came out correct, $b_0 = 1$, $b_3 = 0$, implying that despite the fluctuations of its simplexes, 
the cell assembly complex $\mathcal{F}_{W}$ does not disintegrate into pieces and remains contractible in higher ($D > 2$) 
dimensions, i.e., that the topological fluctuations in the hippocampal map are bounded to $1D$ paths and $2D$ surfaces.
%%%%%%%%%%%%%%%%%%%%%%%%%%%%%%%%%%%
%%%%%%%%%%%%%%%%%%%%%%%%%%%%%%%%%%%

Further numerical analyses demonstrate that, as the memory window increases, the Betti numbers $b_1$ and $b_2$ become 
more stable. In contrast, as the memory window shrinks, the fluctuations of the topological loops intensify (Fig.~\ref{Betti}B). 
This implies that if the characteristic timescale of the network's transience is sufficiently large, then the corresponding coactivity 
complex remains fixed and topologically equivalent to the ``perennial'' cell assembly complex, $\mathcal{F}_{W=\infty} = 
\mathcal{T}_{CA}$. On the other hand, if the simplexes are too unstable, then the cell assembly complex $\mathcal{T}_{CA}$ 
fails to acquire the correct physical structure, no matter how long is the learning  period.

\section{Discussion}
\label{section:discussion}

In physical literature, fluctuating simplicial complexes have been previously used in the context of Simplicial Quantum Gravity 
theories for discretizing quantum fluctuations of the space-time \cite{Ambjorn,Hamber}. A natural requirement for these theories 
is that the quantum fluctuations at the micro-scale should average out in the thermodynamic limit, yielding a smooth space-time 
continuum at the macroscale.

There are certain parallels with cognitive representations of space that emerge from the spiking activity of neuronal ensembles. 
At the micro level, the encoded spatial information is naturally discrete at the cellular level and allows a schematic representation 
in terms of simplicial complexes \cite{PLoS,Arai,Babichev,Schemas}. Moreover, since the various place cell assemblies that spike 
at different locations are transient structures, these auxiliary simplicial complexes fluctuate. 
On the macro level, the emergent hippocampal maps provide a stable topological representation of the physical environment over 
long periods \cite{Meck,Clayton,Brown,eLife}, enabling topological reasoning during animals' spatial navigation \cite{Alvernhe,Poucet}.

Our model provides one explanation for how these two experimentally established aspects of hippocampal neurophysiology can 
coexist. According to the model, while the simplexes of the flickering simplicial complex $\mathcal{F}_{W}$ fluctuate at the 
physiological timescale, its Betti numbers can keep their physical values indefinitely. The fact that the simplex fluctuations can 
average out at the scale of the entire complex suggests that rapid rewiring of the cell assemblies can preserve the stability of 
the cognitive map as a whole.

Although these results were obtained using a simple computational model, we hypothesize that the observed effect reflects a 
more general phenomenon that might apply to physiological mechanisms of synaptic and structural plasticity in the hippocampal 
network.

\section{Methods}
\label{section:methods}

Place cell spiking activity is modeled as a stationary temporal Poisson process with a spatially localized rate \cite{Barbieri} 
that is characterized by the spatial location of the peak, $r_c$, the peak firing amplitude, $f_c$, and the place field size, 
$s_c$. We simulated an ensemble of $N_c = 300$ place cells with log-normally distributed peak firing amplitudes, with the 
mean $\langle f_c \rangle = 14$ Hz, log-normally distributed place field sizes with the mean $\langle s_c \rangle = 17$ cm 
(for details see Methods in \cite{PLoS}).  Spiking is modulated by the 8 Hz $\theta$-oscillations---the basic rhythm of the 
extracellular field in the hippocampus \cite{Buzsaki1}. Neuronal coactivity is defined as firing that occurred over two 
consecutive $\theta$ cycles---a computationally established optimal value \cite{Arai}, which corresponds to the parameters 
estimated from experimental data \cite{Buzsaki1,Buzsaki2}. The topological analyses were implemented using the JPlex 
package \cite{JPlex}.

\section{Acknowledgments}
\label{section:acknow}

The work was supported by the NSF 1422438 grant and by Houston Bioinformatics Endowment Fund.

%\newpage

\end{document}